\documentclass[a4paper]{jpconf}
\usepackage{graphicx}
\usepackage{makeidx}
\usepackage{epsfig}
\usepackage{mathrsfs}
\usepackage{latexsym}
\usepackage{amsfonts}
\usepackage{color}
\usepackage{xcolor}
\usepackage{amssymb}
\usepackage{amsmath}
\usepackage{amsthm}
\usepackage{caption}
\begin{document}
\title{Relation between non trivial M2-branes and D2-branes with fluxes}

\author{M P Garcia del Moral$^{1,a}$, C Las Heras$^{1,b}$}

\address{$^1$ Departamento de F\'isica, Universidad de Antofagasta, Antofagasta, Chile.}



\ead{$^a$maria.garciadelmoral@ua.cl, $^b$camilo.lasheras@ua.cl}

\begin{abstract}
 We show the relation between three non trivial sectors of M2-brane theory formulated in the LCG connected among them by canonical transformations. These sectors correspond to the supermembrane theory formulated on a $M_9\times T^2$ on three different constant three-form backgrounds: M2-brane with constant $C_{-}$, M2-brane with constant $C_{\pm}$ and M2-brane with a generic  constant $C_3$ denoted as CM2-brane. The first two exhibit a purely discrete supersymmetric spectrum once the central charge condition, or equivalently, the corresponding flux condition has been turned on. The CM2-brane is conjectured to share this spectral property once that fluxes $C_{\pm}$ are turned on. As shown in \cite{mpgm14} they are duals to three inequivalent sectors of the D2-branes with specific worldvolume and background RR and NSNS quantization conditions on each case. 

\end{abstract}

\section{Introduction}
Wrapped M2-brane theories were first considered in \cite{Kallosh,Russo3} as potential interesting well-defined quantum sectors of M-theory. In  \cite{deWit3,deWit4} it was shown that compactification by itself does not remove classical instabilities and the spectrum at quantum level remains continuous. However, a sector of the wrapped  M2-brane subject to a topological condition associated to an irreducible wrapping was introduced in \cite{Restuccia} and denoted by supermembrane with central charges.  In \cite{Boulton} the discreteness of its supersymmetric spectrum was proved. Besides the deep characterization of this sector of the theory that describes some of the microscopical degrees of freedom of M-theory, another nontrivial sector was found in \cite{mpgm6} associated to a wrapped M2-brane on the same flat metric and under the presence of $C_{\pm}$ fluxes.  For the particular case when only the $C_{-}$ flux is present both theories are duals or equivalent. In \cite{mpgm14} it was obtained the LCG toroidally wrapped supermembrane with a general constant three form background formulation, named CM2-branes. This sector becomes also nontrivial  under a $C_{\pm}$ flux condition. We will briefly review these sectors, and make some comments on their principal features and show their explicit relations among  them. Moreover, in \cite{mpgm14} we obtained for each sector its D2-brane duals which correspond to D2 branes with specific worldvolume and background RR and NSNS  fluxes. We denote these sectors as nontrivial D2-branes and we will discuss the relation among them. 
\section{Non trivial M2 branes}
Let us consider the Light-Cone (LC) formulation of a M2-brane theory on $M_9\times T^2$ on a constant three-form background $C_{3}$ denoted CM2 \cite{mpgm14}. We may use the residual freedom of the gauge transformations associated to the three-form to set $C_{+-a}=0$. We will choose a constant background with non trivial components given by $C_{\pm ab}$ and $C_{abc}$. We will consider a foliation of M2-brane worldvolume, such that, $\Sigma$ is a Riemann surface of genus one related to the spatial directions. Moreover, we will consider a non trivial flux condition on $T^2$  which implies through its pullback a flux condition on $\Sigma$ as shown in \cite{mpgm14,mpgm6}
\begin{eqnarray}
\int_{T^2}\widetilde{F}_{\pm}
 = k_{\pm} \rightarrow \int_{\Sigma} C_{\pm}= k_{\pm}, 
\end{eqnarray}
where $\widetilde{F}_{\pm} = \frac{1}{2}C_{\pm rs}M^r_pM^s_q d\widetilde{X}^p\wedge d\widetilde{X}^q$ and $C_{\pm} = \frac{1}{2}C_{\pm rs} dX^r\wedge dX^s$, being $\widetilde{X}^r$ the target torus coordinates which are identified with the minimal maps $\widetilde{X}^r=\widehat{X}^r(\sigma^1,\sigma^2)$ \cite{mpgm10}. Due to the nontrivial worldvolume flux condition, the closed one-forms are decomposed $dX^r=M^r_s d\widehat{X}^s+dA^r$, with $dA^r$ a dynamical exact one-form and $dX_h^r$ its harmonic counterpart, being $dX_h^r=M^r_sd\widehat{X}^s$ and $M^1_s+iM^2_s=2\pi R(l_s+m_s\tau)$. The embedding maps $dX^r$ of the compact sector satisfy the  standard wrapping condition $\oint_{C_{S}} dX = 2\pi R(l_s+m_s\tau)$ with $X=X^1+iX^2$.

It is worth to notice that when $C_{\pm rs}=\epsilon_{rs}$, the flux condition induced on $\Sigma$ is in one to one correspondence with the the so-called central charge condition \cite{Restuccia}
\begin{eqnarray}\label{central charge}
\int_{\Sigma} dX^r\wedge dX^s = \epsilon^{rs}det(W) , \quad det(W)=n\neq 0,
\end{eqnarray}
with $W$ the winding matrix.  Therefore, CM2-brane supersymmetric LCG Hamiltonian, as a generalization of the bosonic one obtained in \cite{mpgm14}, is given by 
\begin{eqnarray}\label{HCM2}
	 H_{CM_2}&=&T\int_\Sigma \sqrt{W} d^2 \sigma \left\lbrace \frac{1}{2}  \left(\frac{P_m-C_m^{(2)}}{\sqrt{W}}\right)^2 + \frac{1}{2}  \left(\frac{P_r-C_r^{(2)}}{\sqrt{W}}\right)^2 + \frac{1}{4} \left\lbrace X^m,X^n \right\rbrace^2 + \frac{1}{2}\left( \mathcal{D}_r X^m\right)^2\right. , \nonumber \\
&+& \left. \frac{1}{2}(*\widehat{F})^2 +\frac{1}{4}(\mathbb{F}^{rs})^2 - \bar{\theta}\Gamma^-\Gamma_r\mathcal{D}_r\theta - \bar{\theta}\Gamma^-\Gamma_m\left\{X^m,\theta\right\}\right\rbrace - \int_\Sigma d^2\sigma C_{+},
	\end{eqnarray}	
	subject to the residual symmetry associated to the local and global area preserving diffeomorphisms (APD) constraints,
\begin{eqnarray}
\epsilon^{uv}\partial_u\left[ \frac{P_m \partial_v X^m}{\sqrt{W}} + \frac{P_r \partial_vX^r}{\sqrt{W}}  + \frac{\bar{S}\partial_v\theta}{\sqrt{W}} \right]\approx 0 \label{localM2},\quad
 \oint_{C_S}\left[\frac{P_m dX^m}{\sqrt{W}} + \frac{P_r dX^r}{\sqrt{W}} +  \frac{\bar{S}d\theta}{\sqrt{W}}  \right]\approx 0, \label{globalM2}
\end{eqnarray}
where $m=3,\dots,9$ are indices related to the non-compact sector, $r,s=1,2$ to the compact one and $\theta$ is a Majorana spinor of 32 components.  The background terms are
\begin{eqnarray}
C_m^{(2)} &=& \frac{1}{2}\epsilon^{uv}\partial_u X^{\bar{n}} \partial_v X^n C_{m\bar{n}n} + \epsilon^{uv}\partial_u X^n \partial_v X^r C_{mnr}+ \frac{1}{2}\epsilon^{uv}\partial_u X^r \partial_v X^s C_{mrs} ,  \\
C_r^{(2)} &=& \frac{1}{2}\epsilon^{uv}\partial_u X^m \partial_v X^n C_{rmn} + \epsilon^{uv}\partial_u X^m \partial_v X^s C_{rms},  \\
C_{\pm} &=& \frac{1}{2}\epsilon^{uv}\partial_u X^m \partial_v X^n C_{\pm mn}+  \epsilon^{uv}\partial_u X^n \partial_v X^r C_{\pm mr}+ \frac{1}{2}\epsilon^{uv}\partial_u X^r \partial_v X^s C_{+rs},
\end{eqnarray}
The gauge symplectic curvature and the Hodge dual of the flux curvature are respectively given by
\begin{eqnarray}
\mathbb{F}_{rs} = D_r\mathcal{A}_s - D_s\mathcal{A}_r + \left\lbrace \mathcal{A}_r, \mathcal{A}_s \right\rbrace , \quad
*\widehat{F} = \frac{\epsilon^{uv} \widehat {F}_{uv}}{2\sqrt{W}}= \frac{1}{2}\epsilon_{rs}\left\lbrace X_h^r,X_h^s\right\rbrace, 
\end{eqnarray}
with $\mathcal{D}_r \cdot = D_r \cdot  + \left\lbrace \mathcal{A}_r, \cdot \right\rbrace$.
In order to relate the CM2-brane with  the M2-brane with $C_{\pm}$ fluxes described in (\ref{HCM2}), we find as a new result that there exists a canonical transformation 
\begin{eqnarray}
\widehat{P}_m=P_m-C_m^{(2)} , \label{canonical1} \quad
\widehat{P}_r=P_r-C_r^{(2)}, \label{canonical2}
\end{eqnarray}
that preserves all brackets of the theory and the kinematic term
\begin{eqnarray}
\int_{\Sigma} \left( \widehat{P}_m \Dot{X}^m + \widehat{P}_r \Dot{X}^r+\bar{S}\dot{\theta}\right) = \int_{\Sigma} \left( P_m \Dot{X}^m + P_r \Dot{X}^r+\bar{S}\dot{\theta}\right).
\end{eqnarray}
In fact, it can be checked that this relation holds before imposing the flux conditions. The resultant Hamiltonian is given by
\begin{eqnarray}\label{Hflux}
	 H_{C_{\pm}}&=&\int_\Sigma \sqrt{W} d^2 \sigma \left\lbrace \frac{1}{2}  \left(\frac{\widehat{P}_m}{\sqrt{W}}\right)^2 + \frac{1}{2}  \left(\frac{\widehat{P}_r}{\sqrt{W}}\right)^2 + \frac{1}{4} \left\lbrace X^m,X^n \right\rbrace^2 + \frac{1}{2}\left( \mathcal{D}_r X^m\right)^2\right. , \nonumber \\
&+& \left. \frac{1}{2}(*\widehat{F})^2 +\frac{1}{4}(\mathbb{F}^{rs})^2 - \bar{\theta}\Gamma^-\Gamma_r\mathcal{D}_r\theta - \bar{\theta}\Gamma^-\Gamma_m\left\{X^m,\theta\right\}\right\rbrace - \int_\Sigma d^2\sigma C_{+},
	\end{eqnarray}	
subject to (\ref{localM2}). Therefore, the non trivial CM2-brane  Hamiltonian \cite{mpgm14} is equivalent through a canonical transformation to the M2 brane with $C_{\pm}$ fluxes. If a matrix regularization is provided, the spectrum must share the same discreteness properties. One can realize that there is not a flux condition on the constant $C_{abc}$ three-form, however it is only due to the particularity of the background considered. For more general toroidal backgrounds, an analogous flux condition should be imposed. On the other hand in \cite{mpgm6} the authors showed that the M2-brane with $C_{\pm}$ fluxes is equivalent (or dual) to the M2-brane with central charge \cite{Ovalle1} when we set a background with $C_{+rs}=0$ and a flux condition over $C_-$ is imposed. One could also consider that only a quantized  constant $C_{+rs}$ is present -by imposing $C_{-rs}=0$ through a gauge fixing-, in which case the theory corresponds to the M2-brane with central charge but with a constant shift in the Hamiltonian and on its spectrum \cite{mpgm6}.

In consequence, there are at least three non trivial sectors of the toroidally compactified M2-brane on a flat superspace and constant three form background with good quantum properties, i.e. the discreteness of their spectrum, related among them by canonical transformations. In the next section we will discuss their non trivial D2-brane duals.


\section{Non Trivial D2-branes}
Let us consider the Dirac Born Infeld  (DBI) LC formulation of a D2-brane coupled to a constant RR and NSNS flux background. The LCG formulation without the coupling to the background fields, was known from the works of  \cite{Manvelyan1,Manvelyan2,Lee}. We generalized this result in \cite{mpgm14} by considering a D2-brane on $M_8\times T^2$ on a constant $C_{3}$ background. The physical Hamiltonian was obtained after a proper elimination of the non physical degrees of freedom through a canonical transformation. In fact, as the coupling with  background three-form in eleven and ten dimensions, respectively, is given by, essentially, a Wess Zumino term, a similar structure arise when the LCG formulation is considered. Therefore, the previous M2-brane LCG Hamiltonian formulation experience on $M_9\times T^2$ helps to obtain the corresponding one associated to a D2-brane coupled to RR and NSNS background fields
\begin{eqnarray}
\mathcal{H}\label{HCD2}
&=& \frac{1}{2\sqrt{W}}\left[(P_M+B_M) + \Pi^u\Pi^v\gamma_{uv} + G\right] - C^{(10)}_+ - B_+
\end{eqnarray}
where $G=\gamma+\mathcal{F}$ with $\mathcal{F}=det(F_{uv} + B_{uv})$. Moreover $B_M=\Pi^u\partial_u X^N B_{MN}$ and $B_+=\Pi^u\partial_u X^N B_{+N}$ as in \cite{mpgm14}. This Hamiltonian is subject to: a residual constraint $\phi$ related with the D2-brane APD and a first class Gauss constraint $\chi$ associated to the BI U(1) symmetry over the D2 worldvolume $\widetilde{\Sigma}$,
\begin{eqnarray}
    \phi=\epsilon^{uv}\partial_u\left[\frac{P_M\partial_vX^M}{\sqrt{W}}+\frac{\Pi^uF_{vu}}{\sqrt{W}}  \right]  ; \quad
    \chi =\partial_u \Pi^u , 
    \end{eqnarray}
where $M=1,\dots,8$ are the transverse indices. We have shown in \cite{mpgm14} that this Hamiltonian may be understood as the dual of CM2 brane theory on $M_{10}\times S^1$. If a toroidally compactified background target space, like $M_8\times T^2$, is considered and a quantization condition on RR and NSNS background fields is imposed, they imply flux conditions on $T^2$ and $\widetilde{\Sigma}$. In order to see this, let us discuss the next scenarios.
\subsection{Non trivial D2-branes on a constant RR and NSNS flux background}
Let us consider the LCG formulation of D2-branes on $M_8\times T^2$ coupled to RR and NSNS background fields, in such a way that $C^{(10)}_{+-M}$, $B_{+-}$ and $B_{-M}$ has been set to zero fixing the  RR three-form and NSNS two-form gauge invariance, respectively. Moreover, the only non trivial but constant components that we will consider by fixing the background are $C_{\pm rs}$, $B_{rs}$ and $B_{+r}$ with $r=1,2$. 

Let us consider quantization conditions on two well-defined two forms $\widetilde{F}_{\pm}$ and $\widetilde{B}_2$ as
\begin{eqnarray}
\int_{T^2} \widetilde{F}_{\pm}=k_{\pm} \quad,\quad \int_{T^2}\widetilde{B}_2=k_B
\end{eqnarray}
where $k_B,k_{\pm}\in\mathbb{Z}/\{0\}$, $\widetilde{F}_{\pm} = \frac{1}{2}C^{(10)}_{\pm rs}M^r_p M^s_q d\widetilde{X}^p \wedge d\widetilde{X}^q$, $\widetilde{B}_2=\frac{1}{2}B_{rs} M^r_p M^s_q d\widetilde{X}^p\wedge d\widetilde{X}^q$, $B_{rs}=b\epsilon_{rs}$ and $C^{(10)}_{\pm rs}$ being defined as in \cite{mpgm14}. It can be checked that these flux conditions on $T^2$ implies the following D2-brane worldvolume flux conditions
\begin{eqnarray}\label{quantizationwv}
\int_\Sigma C^{(10)}_{\pm}=k_{\pm} \quad,\quad \int_{\widetilde{\Sigma}} B_2 =k_B,
\end{eqnarray}
with $C^{(10)}_{\pm}=\frac{1}{2}C^{(10)}_{\pm rs}dX^r\wedge dX^s$ and $B_2=\frac{1}{2} B_{rs} d X^r\wedge d X^s$ where, as before, the one-forms decomposed on its exact and harmonic part, and we have also considered an identification $\widetilde{X}=\widehat{X}(\sigma^1,\sigma^2)$.
The LCG Hamiltonian of a D2-brane on $M_8\times T^2$ on a constant RR and NSNS background in \cite{mpgm14} was shown to be
\begin{eqnarray}\label{HCMIMD2}
H_{D2}&=& \int d^2\sigma \left\lbrace \frac{1}{2}\frac{(P_\alpha)^2 }{\sqrt{W}} +\frac{1}{2}\frac{(P_r- B_r)^2}{\sqrt{W}}+ \frac{1}{2}\frac{\Pi^u\Pi^v \gamma_{uv}}{\sqrt{W}}+\frac{1}{2}\frac{\widetilde{G}}{\sqrt{W}} \right., \nonumber \\
&+&\left.
\frac{1}{2}\sqrt{W}\left(\mathcal{D}_r X^\alpha\right)^2+\sqrt{W}\left[ \frac{1}{2}(*\widehat{F})^2  + \frac{1}{4}(\mathbb{F}^{rs})^2\right]- C^{(10)}_{+} - B_{+} \right\rbrace
\end{eqnarray}
subject to the residual constraints associated to Gauss law and the local and global APD
\begin{eqnarray}
\partial_u\Pi^u &\approx& 0  \label{Gauss2}\\
\epsilon^{uv}\partial_u\left[ \frac{P_\alpha \partial_v X^\alpha}{\sqrt{W}} + \frac{P_r \partial_vX^r}{\sqrt{W}} + \frac{\Pi^w F_{vw}}{\sqrt{W}} \right] &\approx& 0 \label{D2local2}\\
 \oint_{C_S}\left[\frac{P_\alpha \partial_v X^\alpha}{\sqrt{W}} + \frac{P_r \partial_vX^r}{\sqrt{W}} + \frac{\Pi^w F_{vw}}{\sqrt{W}}\right] d^v\sigma &\approx& 0 \label{D2global2}
\end{eqnarray}
where $\widetilde{G}=\widetilde{\gamma} + \mathcal{F}$, $\widetilde{\gamma}=det(\partial_u X^\alpha \partial_v X_\alpha )$, $\mathcal{F}=det\left[ F_{uv} + \sqrt{W}k_B\epsilon_{uv}\left(*\widehat{F} + \frac{1}{2}\epsilon_{rs}\mathbb{F}^{rs}\right)\right]$ and $F=dA$.
 We have shown in \cite{mpgm14} the Hamiltonian (\ref{HCMIMD2}) is dual in the sense of \cite{Duff5,Townsend4} to CM2 brane theory on a constant flux background (\ref{HCM2}) by a proper fixing of the corresponding background. 

We can mention that the worldvolume flux condition can also be interpreted in terms of  the irreducible wrapping condition -found in \cite{Restuccia} in the context of wrapped M2-branes- of D2-branes \cite{mpgm14}. It is clear that if $C_{\pm rs}=\epsilon_{rs}$ and $B_{rs}=\epsilon_{rs}$, then both quantization conditions (\ref{quantizationwv}) on $\widetilde{\Sigma}$ are in one to one correspondence with this last one.
\subsection{Non trivial D2-branes on a constant RR background}
 Let us consider once again the LCG formulation of a D2-brane on $M_8\times T^2$, where $C^{(10)}_{+-M}= B_{+-}=B_{-M}=0$ have been set to zero using RR three-form and NSNS two-form gauge invariance and the only non trivial but constant components that we will consider by fixing the background are the $C_{\pm rs}$ and $B_{+r}$ with $r=1,2$, ($B_{MN}=0$ on expression (\ref{HCD2})). Therefore, we may consider the flux condition over $\widetilde{F}_{\pm}$ on $T^2$ which implies a flux condition on $C_{\pm}^{(10)}$ over the worldvolume. In this case we obtained in \cite{mpgm14}
\begin{eqnarray}\label{HD2fluxx}
H&=& \int d^2\sigma \left\lbrace \frac{1}{2}\frac{(P_\alpha)^2 }{\sqrt{W}} +\frac{1}{2}\frac{(P_r)^2}{\sqrt{W}}+ \frac{1}{2}\frac{\Pi^u\Pi^v \gamma_{uv}}{\sqrt{W}}+\frac{1}{2}\frac{\widetilde{G}_{DBI}}{\sqrt{W}} \right., \nonumber \\
&+&\left.
\frac{1}{2}\sqrt{W}\left(\mathcal{D}_r X^\alpha\right)^2+\sqrt{W}\left[ \frac{1}{2}(*\widehat{F})^2  + \frac{1}{4}(\mathbb{F}^{rs})^2\right]- C^{(10)}_{+} - B_{+} \right\rbrace
\end{eqnarray}
subject to (\ref{Gauss2}), (\ref{D2local2}) and (\ref{D2global2})
where $\widetilde{G}_{DBI}=\widetilde{\gamma} + F$. As shown in \cite{mpgm14}, this LCG Hamiltonian for a D2-brane on $M_8\times T^2$ with $C^{(10)}_{\pm}$ fluxes, is dual to the Hamiltonian (\ref{Hflux}) which corresponds to M2-brane with $C_{\pm}$ fluxes \cite{mpgm6} and a NSNS field contribution. However, it is worth to mention that the fact that there are canonical transformations (\ref{canonical1}) on $D=11$, which allow us to understand CM2 brane with $C_{\pm}$ fluxes \cite{mpgm14} as equivalent to M2-brane with fluxes $C_{\pm}$ \cite{mpgm6} and M2-brane with central charges \cite{Restuccia}, does not implies the same relation in $D=10$. However, one can relate (\ref{HCMIMD2}) with (\ref{HD2fluxx}), by switching off the transverse components of the NSNS background field $B_{MN}=0$, before the imposition of fluxes. They correspond to different D2-brane sectors depending on the the transverse components of NSNS background field.

\subsection{Non trivial D2-branes with an irreducible wrapping condition} We consider the dual of a M2-brane LCG Hamiltonian with a $C_-$ flux condition, which is given by (\ref{Hflux}) setting $C_+=0$. It is an equivalent topological condition to the central charge condition in the M2-brane theory \cite{Restuccia}. Its dual is the standard toroidally wrapped D2-brane subject to  a non trivial flux condition on $T^2$ associated to the $C_{-rs}^{(10)}=\epsilon_{rs}$. This condition can also be seen as a topological irreducible wrapping condition on the D2-brane on a flat toroidal background. As shown in \cite{mpgm14} its D2-brane action is dual to M2-brane with central charges. 

\subsection{Non trivial D2-branes on a constant NSNS background}
We may also consider the D2-branes on $M_8\times T^2$ only coupled to constant NSNS background fields in such a way that $B_{+-}=B_{\pm r}=0$ fixing the residual gauge transformation of the two-form. Moreover, we also impose $C_{\pm}^{(10)}=0$ by fixing the background, such that the only non trivial components considered are given by $B_{rs}=b\epsilon_{rs}$. By imposing a quantization condition on $\widetilde{B}_2$ over $T^2$, which for a constant 2-form is equivalent to a 2-form flux condition  over $\widetilde{\Sigma}$,  it can be checked that the Hamiltonian is given by (\ref{HCMIMD2}) with $C_{+}=B_{+}=0$.  The B-field quantization condition is also equivalent to the D2-brane irreducible charge condition on the D2 worldvolume for $b=1$, however, the Hamiltonians differ. 
\begin{figure}
\centering
    \includegraphics[scale=0.5]{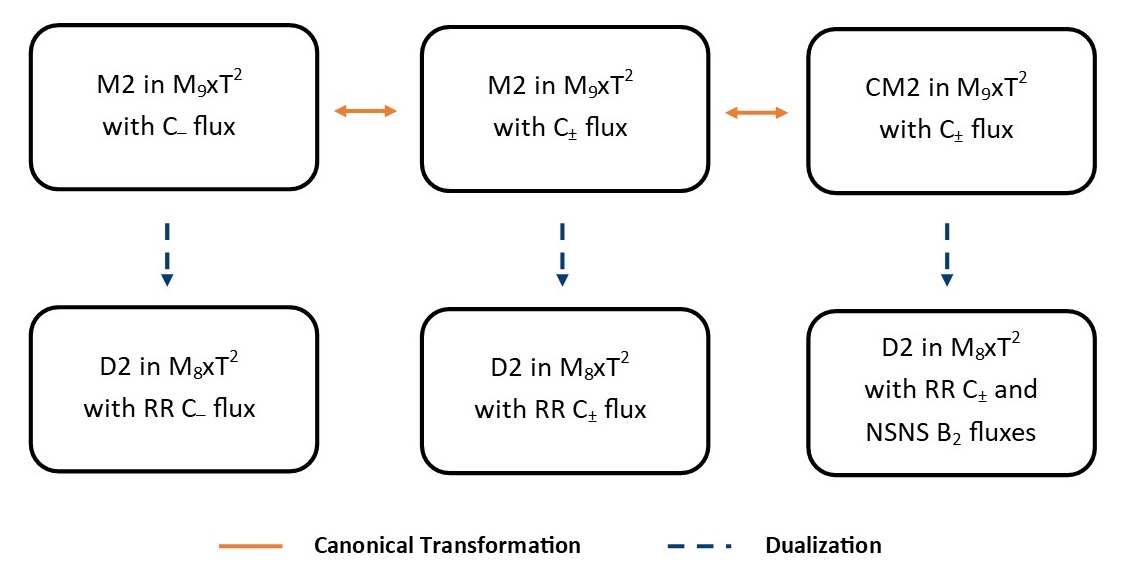}
    \caption{Relations between the non trivial M2-brane and D2-branes with RR and NSNS fluxes}
    \label{fig:my_label}
\end{figure}

\section{Discussion} We have shown that the three non trivial sectors of M2-brane theory with discrete spectrum are in fact related to each other by canonical transformations. These sectors named as: M2-brane with central charge \cite{Restuccia,Ovalle1} (or equivalently, M2-brane with $C_{-}$ fluxes), M2-brane with $C_{\pm}$ fluxes \cite{mpgm6} and CM2-brane with $C_{\pm}$ fluxes \cite{mpgm14} are duals to non trivial sectors of the D2-branes formulated on the same target space $M_8\times T^2$ with specific RR and NSNS flux content for each case. 
These cases are: 1) The CM2-brane with $C_{\pm}$ fluxes dual to a D2-brane with RR and NSNS fluxes on the target and on the worldvolume. 2) The M2-brane with $C_{\pm}$ fluxes dual to a D2-brane with RR flux on the target but vanishing transverse components of the B-field on the worldvolume and 3) the M2-brane with $C_{-}$ fluxes dual to a D2-brane with $C_-^{(10)}$ flux (also equivalent to an irreducible wrapping condition).  The non trivial D2-brane sectors can be related one to another by switching off some of the background fields: i.e. from 1) to 2) imposing $B_{rs}=0$ and from 2) to 3) $C_{+}^{(10)}=B_{\pm r}=0.$ We may also consider a specific background with a flux condition over the only non trivial but constant background field $B_{rs}$. This is a particular case of 1) with the D2-brane dual of CM2-brane with $C_-$ fluxes. In summary, the discreteness condition \cite{Boulton} over the M2-branes imply non-vanishing worldvolume and background fluxes over its D2-brane duals. 

\section*{Acknowledgements} 
CLH is supported by CONICYT PFCHA/DOCTORADO BECAS CHILE/2019- 21190263 and by  ANT1956 project of Antofagasta U. The authors also thanks to Semillero funding SEM18-02 from Antofagasta U. and to the international ICTP Network NT08 for kind support. 

\section*{References}

\providecommand{\newblock}{}

\end{document}